\documentclass[manuscript=article]{achemso}
\usepackage{graphicx}
\usepackage[table]{xcolor}
\usepackage[breaklinks=true,colorlinks=true,linkcolor=blue,urlcolor=blue,citecolor=blue]{hyperref}
\usepackage{amsmath}    
\usepackage{bm}
\usepackage{textcomp}
\usepackage{mathtools}
\usepackage{cleveref}
\usepackage{flexisym}
\usepackage{braket}     
\usepackage{graphicx}   
\usepackage{verbatim}   
\usepackage{color}      
\usepackage{subfigure}  
\usepackage{hyperref}   
\usepackage[utf8]{inputenc}
\usepackage[english]{babel}

\usepackage{pdfpages} 

\usepackage{xr-hyper} 
\newcommand{\HH}{\mathcal{H}}

\title{Higher-order exchange interactions in two-dimensional magnets}
\author{Alexey Kartsev}
\affiliation{School of Mathematics and Physics Queen's University Belfast Belfast, BT7 1NN, United Kingdom}
\author{Mathias Augustin}
\affiliation{School of Mathematics and Physics Queen's University Belfast Belfast, BT7 1NN, United Kingdom}
\author{Richard F. L. Evans}
\affiliation{Department of Physics, The University of York, YO10 5DD, United Kingdom}
\author{Kostya S. Novoselov}
\affiliation{Department of Material Science \& Engineering, National University of Singapore, Block EA, 
9 Engineering Drive 1, 117575, Singapore} 
\author{Elton J. G. Santos}
\affiliation{Institute for Condensed Matter Physics and Complex Systems, School of Physics and Astronomy, The University Edinburgh, EH9 3FD, United Kingdom}
\email{esantos@ed.ac.uk}

\date{\today}


\begin{document}

\maketitle

\begin{abstract}
{\bf Magnetism in recently discovered van der Waals materials 
has opened new avenues in the study of fundamental spin interactions in truly two-dimensions. 
A paramount question is what effect higher-order interactions
beyond bilinear Heisenberg exchange have on the magnetic properties of few-atom 
thick compounds. 
Here we demonstrate that biquadratic exchange interactions, 
which is the simplest and most 
natural form of non-Heisenberg coupling, assume a key role 
in the magnetic properties of layered magnets. 
Using a combination of nonperturbative analytical techniques, non-collinear 
first-principles methods and classical 
Monte Carlo calculations that incorporate higher-order exchange, 
we show that several quantities including magnetic anisotropies, 
spin-wave gaps and topological spin-excitations are intrinsically 
renormalized leading to further 
thermal stability of the layers. 
We develop a spin Hamiltonian that also contains antisymmetric exchanges 
(e.g. Dzyaloshinskii–Moriya interactions) to successfully rationalize 
numerous observations currently under debate,  
such as the non-Ising character of several compounds despite a strong 
magnetic anisotropy, peculiarities of the magnon spectrum 
of 2D magnets, and the discrepancy between measured 
and calculated Curie temperatures. Our results lay the foundation 
of a universal higher-order exchange 
theory for novel 2D magnetic design strategies. 
}
\end{abstract}

\maketitle

\section{Main}

The finding of magnetism in atomically thin van der Waals (vdW) 
materials\cite{firstCrI3,CrGeTe}  
has attracted an increasing amount of 
interest in the investigation of novel magnetic phenomena at the nanoscale\cite{Guguchiaeaat3672}. 
Important in the understanding and applications of 2D vdW magnets in real 
technologies is the elucidation 
of fundamental interactions that determine their magnetic properties, 
such as the exchange interactions between the spins. 
These interactions govern the magnetic ordering and can be either symmetric,
which determine collinear ferromagnets and anti-ferromagnets, 
or antisymmetric, which promote topological non-trivial spin textures, e.g. skyrmions. 
Higher-order exchange terms involving the hopping 
of two or more electrons play a pivotal role in the 
spin-ordering of low-dimensional nanostructures.  
For instance, biquadratic (BQ) exchange interactions are critical in the 
elucidation of the magnetic features of several systems, such as multilayer 
materials\cite{Slonczewski91}, perovskites\cite{Nicola15}, 
iron-based superconductors\cite{Wysocki:2011aa,Iron-BQ}, 
iron tellurides\cite{Ashvin09} and oxides\cite{Harris63}.  
Indeed, in materials where the exchange is for 
some reason weak\cite{Nagaev_1982} (e.g. low-temperature magnets)
BQ exchange has a particularly strong influence being the case of several 2D magnets
discovered so far\cite{Gongeaav4450,Novoselov19}.

Here we identify that several families of 2D magnets including metals, 
insulators and small band gap semiconductors, develop substantially 
large BQ exchange interactions. The delicate interplay between 
the superexchange process through the non-magnetic atom and the 
Coulomb repulsion at neighboring spin-sites involving more than one 
electron induces sizable corrections to the total energy of the systems beyond bilinear spin 
models, e.g. Kitaev, Heisenberg, Ising. We developed a generalized non-Heisenberg 
spin Hamiltonian that includes both BQ and Dzyaloshinskii–Moriya interactions (DMI), 
providing a universal picture of the spin properties of 2D vdW 
magnets. We show that while BQ exchange interactions 
give substantial contributions to the thermal properties, such as in  
critical temperatures (T$_{\rm c}$), non-collinear spin effects 
via DMI are negligible. The model also offers insights via 
analytical equations into the spin excitations of 2D materials as a 
general formalism is developed for materials that hold honeycomb 
crystal structure, ferromagnetism and out-of-plane easy axis in a universal basis. 
Our results provide the conceptual framework to understand a variety of 2D magnetic 
materials in a consistent way explaining numerous experimental observations, including 
controversies on the magnetic properties of CrI$_{3}$ or CrBr$_3$.

\section{Biquadratic exchange interactions in 2D magnets} 


In order to calculate the different exchange contributions to the total energy 
at the level of first-principles methods (see Supplementary Sections \ref{SI-sec:computational}-\ref{SI-BQcalc} for details), 
we map the angular dependence of the 
spins ${\bf S}_{j}=\mu_{s}{\bf s}_{j}$, where $\mu_{s}$ is the magnetic 
moment and $|{\bf s}_{j}|=1$, in the unit cell of the layered material (Fig.~\ref{fig:etot_rotscheme}{\bf a}). 
We rotate the spins by $\theta$ between two known spin configurations: from ferromagnetic (FM) 
at $\theta = 0$, to anti-ferromagnetic (AFM) at $\theta = 180^{\rm o}$.  
Small steps in $\theta$ generate a path of quasi-continuously configurations where both energy 
and magnetization are allowed to relax self-consistently without any fixed constraint on the direction.
The resulting curve in energy includes contributions from bilinear (BL) exchanges up to   
higher-order terms, e.g. BQ exchange interactions. 
We quantify the contribution of each kind of interactions using the following 
non-Heisenberg spin Hamiltonian:  

\begin{equation}
{\cal H} = -\sum_{ij} J_{ij} ({\bf S}_i \cdot {\bf S}_j) - \sum_{ij} \lambda_{ij} S_i^z S_j^z -\sum_i D_i \left({\bf S}_i \cdot {\bf e}_i\right)^{2} - \sum_{ij} K_{ij} \left({\bf S}_i \cdot {\bf S}_j \right)^2 
\label{eq:hamil_bq}
\end{equation} 
where ${J}_{ij}$ and $ \lambda_{ij}$ are the isotropic and anisotropic BL exchanges 
between spins {\bf S}$_i$ and ${\bf S}_j$ on atomic sites $i$ and $j$; 
$D_i$ is the on-site anisotropy with easy axis ${\bf e}_i$; and $K_{ij}$ is the BQ exchange interactions 
which is due to electron hopping between two adjacent sites\cite{Takahashi77}. 
We restrict the discussions to first nearest-neighbor 
BQ interactions, that is, $K_{ij} = K_{bq}$, and $K_{ij} =0$ otherwise. 
This assumption has been shown to be sufficient to study a variety of nanomagnetic 
systems with higher-order exchanges\cite{Nicola15, Wysocki:2011aa, Harris63, Ashvin09}. 
We can write Eq.\ref{eq:hamil_bq} in a similar form separating the terms with 
angular and non-angular dependence as 
$E^{tot}_{bq}\left(\theta \right)=A^{bq}_0+A^{bq}_1\cdot S^2 \cos(\theta)+A^{bq}_2\cdot S^4 \cos^2(\theta)$, 
where $S$ is the spin moment. 
The different coefficients can be interpreted as the corresponding amount 
of BL ($A^{bq}_1$) and BQ ($A^{bq}_2$) exchanges, and the 
on-site energy as the spins are perpendicular to each other ($A^{bq}_0$). 
Supplementary Section \ref{SI-BQcalc} gives a thorough discussion on these coefficients 
and how to extract analytical equations for their interpretation using Eq.\ref{eq:hamil_bq}. 
We extract $A^{bq}_0$, $A^{bq}_1$ and $A^{bq}_2$  
from first-principles simulations for the variation of the total energy 
versus $\theta$ (Supplementary Section \ref{SI-sec:table-fitting}).  
 
We apply this procedure for a total of 50 compounds 
including the most common 2D magnets studied up to date including 
several families of trihalides (MX$_{\rm 3}$, M=Ti, V, Cr, Mn, Fe, Ni, Cu;  X=F, Cl, Br, I), 
metal tribromides (MBr$_{\rm 3}$, M=Mn, Cu, Fe, V),  
chromium based ternary tellurides (Cr$_{\rm 2}$X$_{\rm 2}$Te$_{\rm 6}$, X=Ge, P, Si), 
metal based ternary chalcogenides (M$_{\rm 2}$P$_{\rm 2}$X$_{\rm 6}$, M= V, Cr, Mn, Fe, Co, Ni; X=S, Se, Te), and 
transition metal dichalcogenides (MX$_{\rm 2}$, M=Co, Fe, V; X=S, Se, Te) of different phases ($2H$, $1T$).  
Surprisingly, as the spins are spatially rotated away 
a sizable deviation from a $\cos \theta$-like behavior characteristic of 
BL exchange interactions (shaded area) is 
observed (Fig.~\ref{fig:etot_rotscheme}{\bf b-e}). 
The overall trend seems 
independent of the stoichiometric formula or the atomic elements composing the structure. 
The universal character of the BQ exchange interactions can be appreciated clearer  
in Fig.~\ref{fig:etot_rotscheme}{\bf f} where a quadratic regression (Q-Regression) 
can be undertaken over the computed dataset. 
Materials with alike chemical environment (e.g. bond lengths, electron affinity, binding energy)  
such as CrI$_3$, CrBr$_3$ and CrCl$_3$ present close variation of the energy 
and consequently similar magnitudes of the BQ exchange (Table \ref{table1}). 
Other compounds tend to gain energy with the spin rotation and stabilize in 
a different magnetic coupling, for instance, 
CrF$_3$ (Fig.~\ref{fig:etot_rotscheme}{\bf b}) 
becomes AFM ordered. 
This is also 
the case of Mn-based compounds, 2H-MnS$_2$, 
2H-MnSe$_2$, MnPS$_3$, MnPSe$_3$ (Fig.~\ref{fig:etot_rotscheme}{\bf d-e}), 
which agreed with magnetometry measurements\cite{Kim2019, Makimura93,Corliss59}. 
It is noteworthy that both CuBr$_3$ and 2H-FeS$_2$ have substantially larger  
magnitudes of BQ exchange (Table \ref{table1}) comparable to more complex 
materials, i.e. ferropnictides, where spin and electronic correlations 
are known to play a key role in the determination of their superconducting and strongly-correlated 
properties\cite{BQpictinides}.  This result calls for further theoretical and experimental work on 
these two compounds whether novel electronic interactions can be found. 
Furthermore, we noticed that several other compounds 
could not show a clear trend with $\theta$ apart from those 
where a specific magnetic ordering is stabilized, e.g. $\theta=0,180^{o}$ 
(Supplementary Section \ref{SI-sec:complex-materials}). 
Materials that are unable to stabilize different spin orientations 
are either due to strong magnetic anisotropy where 
a preferential spin orientation is too strong to be tilted (e.g. Ising magnets) 
or different spin solutions are not energetically stable ending up in 
non-magnetic phases\cite{Blugel01}. 

We have also checked whether other models  
can give a sound description of the magnetic properties of the quadratic 
dependence of the energy as a function of $\theta$. 
In particular, we considered two models: a Kitaev model\cite{Kitaev06,Laurent18}, and 
a Heisenberg model including biquadratic on-site 
magnetic anisotropy.  
For the latter, the magnitudes of the 
biquadratic on-site anisotropies are several orders of magnitude smaller than 
BQ exchange within the range $3.87 -14.32~\mu$eV (see details in Supplementary Section \ref{SI-sec:BQ-anisotropy}). 
For the former, there is no quadratic dependence on $\theta$ 
for the Kitaev model. We can expand the Kitaev Hamiltonian\cite{Kitaev06} 
assuming a rotation by angle $\theta$ between the spins to show that 
a fitting equation of the form of $E^{tot}_{Kitaev}\left(\theta \right)=B_0+B_1 \sin(2\theta) $ 
can be extracted (see details in Supplementary Section \ref{SI-sec:kitaev}). 
These results suggest that BL models are insufficient to describe the 
magnetic features of 2D vdW magnets.

\section{A Hubbard-based microscopic model}


To understand the microscopic mechanism of the BQ exchange in 2D materials, we will use the 
following 2D half-filled Hubbard Hamiltonian applied to the honeycomb lattice (Fig.~\ref{fig2}{\bf a-b}): 
\begin{equation}
{H}=-t_{eff} \sum_{<ij>,\sigma} c^{\dagger}_{i,\sigma}c_{j,\sigma}+ U_{} \sum_{i} n_{i\uparrow} n_{\mathrm i\downarrow} + \sum_{i,\sigma}\mu_{i,\sigma}n_{i,\sigma}, 
\label{HubHam}
\end{equation}
where indices $i$ and $j$ denote the lattice sites, the $d$-spin states on the metal atoms (M$_{\rm A,B}$)
are labeled as $\sigma=\uparrow$, $\downarrow$, the sum $<ij>$ is over the nearest 
neighbors and $t_{eff}$ is the effective 
nearest neighbor hoping between M$_{\rm A,B}$ ions. $t_{eff}$ is due to the hybridization 
between $3d^n$ and $np$ orbitals ($n=2,3,4,5$ depending on the atomic element involved) 
at M$_{\rm A,B}$ and X atoms, respectively (Fig.~\ref{fig2}{\bf b}).  
The direct hopping between M$_{\rm A,B}$ scales with $t_{dd}\sim r^{-5}$ and therefore 
for relatively big distances can be neglected\cite{Takahashi77}. 
$U>0$ is the non-negative on-site Coulomb repulsion, 
$c^\dagger_{i,\sigma}$($c_{i,\sigma}$) is the creation (annihilation) operator 
for a fermion with spin $\sigma$ at site $i$, 
$n=c^\dagger_{i,\sigma}c_{i,\sigma}$ is the density operator and $\mu_{i,\sigma}$ is the 
chemical potential which controls the filling of the bands. Second-order perturbation 
theory in Eq.\ref{HubHam}, assuming $U_{} >> t_{eff}$, gives the energy contributions 
of the Heisenberg exchanges\cite{andersonsuperexchnage,Dionne-book}:  
\begin{eqnarray}
J_{\mathrm{FM}}^{(2)}=t_{eff}^2/(U-U_{ex}) \nonumber \\
J_{\mathrm{AFM}}^{(2)}=t_{eff}^2/(U+U_{ex}) \nonumber  \\
J_{\mathrm{bl}}=t_{eff}^2\frac{2U_{ex}}{U^2-U^2_{ex}} 
\label{BL-exchanges}
\end{eqnarray} 
where $J_{\mathrm{FM}}^{(2)}$ and $J_{\mathrm{AFM}}^{(2)}$ are the exchange energies for 
FM and AFM coupling at second-order, and the BL exchange 
is defined as $J_{\mathrm{bl}}= J_{\mathrm{FM}}^{(2)} - J_{\mathrm{AFM}}^{(2)}$. 
$U_{ex}$ is an energy correction for the internal 
spin exchange whether a spin flip is required during the hopping between 
M$_{\rm A,B}$ sites (Fig.~\ref{fig2}{\bf b})\cite{Dionne-book}. 
Such term can be used to stabilize 
or destabilize spin transfer through the M$_{\rm A,B}-$X covalent bonds 
as the exchange occurs\cite{Takahashi77,Dionne-book}. If the electron-hopping 
is to an occupied orbital of the neighboring site, $U_{ex}$ will favor AFM alignment 
of the two spins via a superexchange interaction (see Supplementary Section \ref{SI-sec:fm-afm_competition} for details). 
However, if the electron-hopping is to 
an unoccupied or a virtual state, a FM alignment will be favored by $U_{ex}$\cite{spin_chain2}. 
The competition between the amount of energy $U_{ex}$ to stabilize a specific coupling and 
the Coulomb repulsion $U$ between the electrons at an energy state 
may compensate each other leading to a small value of $J_{\mathrm{bl}}$. 
Indeed, several 2D magnets have shown low-temperature magnetism\cite{firstCrI3,CrGeTe,Novoselov19} 
which is directly related with the small magnitude of the exchange interactions. 
In this case, it is necessary to extend the perturbation in $t_{eff}/U$ to fourth-order in Eq.\ref{HubHam}
involving at least one electron from both M$_{\rm A,B}$ sites which resulted in\cite{Takahashi77}: 
\begin{eqnarray}
J_{\mathrm{FM}}^{(4)} \propto t_{eff}^4/(U-U_{ex})^3 \nonumber \\ 
J_{\mathrm{FM}}^{(4)} \propto t_{eff}^4/(U+U_{ex})^3 \nonumber  \\ 
\label{BQ-exchanges}
K_{\mathrm{bq}}=t_{eff}^4\frac{2(3U^2U_{ex}+U^3_{ex})}{(U^2-U^2_{ex})^3} 
\end{eqnarray}
with $K_{\mathrm{bq}} = J_{\mathrm{FM}}^{(4)} - J_{\mathrm{AFM}}^{(4)}$ being the 
BQ exchange energy. Both Eqs.\ref{BL-exchanges}$-$\ref{BQ-exchanges} show that 
a competition between FM and AFM couplings takes place once the electrons are 
hopping between different spin sites. The stabilization of one or another 
magnetic order is determined by several factors such as the ligand-field 
splitting $\Delta_0$ between $t_{2g}$ and $e_{g}$ states in 
the metal atom in the honeycomb lattice. 
As the filling of both type of states determines the magnitude of the $3d-sp$ hybridization 
between metals and ligands, we can approach $\Delta_0 \approx U-U_{ex}$ being proportional to the 
bandgap of the material\cite{Nagaev_1982}. It has been shown that the role of 
high-order exchange interactions increases on reduction of the bandgap mediated by the non-magnetic atoms
which is related to the ratio of BQ and BL exchanges\cite{andersonsuperexchnage,ANDERSON63}. 
If we divide Eq.\ref{BQ-exchanges} and Eq.\ref{BL-exchanges} and use the 
definition of $\Delta_0$, we can write a direct relationship between the 
exchange interactions and the bandgap for 2D magnets as: 
\begin{equation}
K_{\mathrm{bq}}/J_{\mathrm{bl}}= t_{eff}\frac{4U^2+\Delta_0^2-2U\Delta_0}{\Delta_0^2(2U-\Delta_0)^2} 
\label{BQ/BL}
\end{equation}
This equation can be understood as a direct interplay between the 
Coulomb repulsion and the hopping of electrons between different sites 
subjected to the crystal field or bandgap of the material. We can 
consider two situations in Eq.\ref{BQ/BL}:  
$\Delta_0  \rightarrow 0$ and $\Delta \rightarrow 2U$
which correspond to small and large bandgap materials, respectively. This resulted in:  
\begin{eqnarray}
\label{Delta_0}
K_{\mathrm{bq}}/J_{\mathrm{bl}}\propto t_{eff}\frac{1}{(\Delta_0)^2}\bigg\rvert_{\Delta_0 \rightarrow 0} \\
\label{Delta_2U}
K_{\mathrm{bq}}/J_{\mathrm{bl}}\propto t_{eff}\frac{1}{(2U-\Delta_0)^2}\bigg\rvert_{\Delta_0 \rightarrow 2U} 
\end{eqnarray}
Figure \ref{fig2}{\bf c} shows the variation of $K_{\mathrm{bq}}/J_{\mathrm{bl}}$ 
as a function of $\Delta_0$ for the core of materials displaying BQ exchange interactions. 
Strikingly, both Eqs.\ref{Delta_0}$-$\ref{Delta_2U} correctly describes the overall behavior 
observed in our simulations. Materials with similar bonding environment, for instance, either in terms of 
Cr (Mn) atoms follow an increase (decrease) of $K_{\mathrm{bq}}/J_{\mathrm{bl}}$ with the 
bandgap, respectively. It is worth mentioning that as the compounds tend to a 
AFM spin alignment\cite{spin_chain2,Nagaev_1982}, i.e. CrF$_3$, they increase 
the value of $K_{\mathrm{bq}}/J_{\mathrm{bl}}$ being the 
case within the Cr-based trihalide family (CrX$_3$, X=F, Cl, Br, I).  
There is also an abrupt change in behavior for narrow bandgap materials and 
metals with no dependence on $\Delta_0$ (inset in Fig.~\ref{fig2}{\bf c}). 
These results indicate that one can design the amount of BQ exchange in a 2D magnet
tuning its bandgap, for instance using an electric bias as recently used in CrI$_3$\cite{2LCrI3}.

\section{Thermal effects in 2D magnets}


To verify whether BQ exchange interactions will have any effect on the thermal properties 
of 2D vdW magnets, we have implemented Eq.\ref{eq:hamil_bq} within the Monte Carlo Metropolis algorithm 
with an adaptive move in the open source {\sc Vampire}\cite{Evans14} software package. In the spin model we assume 
a classical spin vector {\bf S}$_i$ on each atomic site $i$. The 
quantization vector for the spin is a local quantity which intrinsically includes 
the effects of local thermal spin fluctuations, magnon processes and spin excitations. 
In our implementation the BQ exchange 
is quite general and can be applied to any pair-wise exchange interaction of arbitrary range.
We also consider an additional term in Eq.\ref{eq:hamil_bq} including the 
local Zeeman field {\bf B} on the magnetic ions with 
a length of the local atomic moment $\mu_i$ arising 
from the BQ exchange interactions as:  
\begin{equation}
{\cal H} = -\sum_{ij} J_{ij} ({\bf S}_i \cdot {\bf S}_j) - \sum_{ij} \lambda_{ij} S_i^z S_j^z -\sum_i D_i \left({\bf S}_i \cdot {\bf e}_i\right)^{2} - \sum_{ij} K_{ij} \left({\bf S}_i \cdot {\bf S}_j \right)^2 - \sum_i \mu_i {\bf S}_i \cdot {\bf B}_i
	    \label{BQ-vamp}
\end{equation}
For atomistic spin dynamics simulations we calculate the effective magnetic 
field ${\bf B}_i$ by taking the first-derivative of Eq.\ref{BQ-vamp} on the 
different spin components:
\begin{eqnarray}
B_x^i &=& -\frac{1}{\mu_i}\frac{\partial \HH_{\mathrm{}}}{\partial S_x} = 2 K_{\mathrm{bq}} S_x^j \left(S_x^i S_x^j + S_y^i S_y^j + S_z^i S_z^j\right) \nonumber \\ 
B_y^i &=& -\frac{1}{\mu_i}\frac{\partial \HH_{\mathrm{}}}{\partial S_y} = 2 K_{\mathrm{bq}} S_y^j \left(S_x^i S_x^j + S_y^i S_y^j + S_z^i S_z^j\right) \nonumber \\ 
B_z^i &=& -\frac{1}{\mu_i}\frac{\partial \HH_{\mathrm{}}}{\partial S_z} = 2 K_{\mathrm{bq}} S_z^j \left(S_x^i S_x^j + S_y^i S_y^j + S_z^i S_z^j\right)\mathrm{.}
\end{eqnarray}
The effective field is then included within the total effective field describing the 
time evolution of each atomic spin using the stochastic Landau-Lifshitz-Gilbert 
equation\cite{Ellis}. We calculate up to third nearest-neighbors 
$\lambda_{ij}$ and $J_{ij}$ in Eq.\ref{eq:hamil_bq} on a representative set of 
2D magnets, e.g. Cr-based trihalide family (Table \ref{table2}).  
Supplementary Section \ref{SI-NNN_exch} gives a 
thorough discussion on the calculation of the exchange parameters. 
Figure \ref{fig3} shows the behavior of the magnetization 
($M/M_0$, where $M_0$ is the saturation magnetization at 0~K) 
and the logarithm of the magnetic susceptibility ($\ln\chi$) as a function of 
temperature T(K) for CrI$_3$, CrBr$_3$ and CrCl$_3$. 
Intriguingly, the inclusion of BQ exchange interactions give sizable thermal effects on both M/M$_0$ and 
$\ln\chi$ for all materials (Supplementary Section \ref{SI-chi-data}). 
By fitting the Monte Carlo 
simulations with ${\rm M(T)=M_0} \left(1-\frac{\rm T}{\rm T_c}\right)^{\beta}$ (where $\beta$ is the critical exponent) 
we can notice that the Curie temperatures T$_{\rm c}$ changes by several Kelvins with the inclusion of 
BQ exchange interactions. The calculated magnitudes of T$_{\rm c}$ for CrI$_3$ and CrBr$_3$ 
approach closely those measured for both compounds with almost no difference (Fig.~\ref{fig3}{\bf a-d}). 
The lack of experimentally measured T$_{\rm c}$ for monolayer CrCl$_3$ unable 
us to make a clear comparison with our simulations.  
The different amount of nearest-neighbors at BL exchanges (from 1st to 3rd)  
also produces substantial effects even though not enough to 
reproduce the experimental values of T$_{\rm c}$ for CrI$_3$ and CrBr$_3$. 
It is worth mentioning that different groups have reported distinct 
magnitudes of T$_{\rm c}$ for CrBr$_3$\cite{Mak19, kostya19,TingYu19}, 
which may be due to different factors such as sample quality, defects, and doping levels. 
We believe that our simulations still provide an accurate picture showing the 
effects of the underlying exchange interactions for these low-dimensional magnets 
at the limit of an ideal, pristine crystal (see Supplementary Section \ref{SI-CrBr3-data}). 
Moreover, several other materials display larger 
values of BQ exchange interactions (Table \ref{table1}) 
indicating that higher-exchange interactions should be taken into account on the 
description of their magnetic properties. 

\section{Enhancement of magnetic stability}
\label{lineartheory}

To understand the intrinsic effect of higher-order exchange interactions in the 
thermal features of 2D magnetic materials and provide a consistent description 
of the underlying spin interactions, we have developed an analytical 
model based on spin-wave theory\cite{Dyson56,Holstein40}. 
We generalized Eq.\ref{eq:hamil_bq} to second order contributions on the magnetic 
anisotropies which resulted in: 
\begin{equation}
\begin{aligned}
{\cal H} = -\sum_{ij} J_{ij} ({\bf S}_i \cdot {\bf S}_j) - \sum_{ij} \lambda_{ij} S_i^z S_j^z  
-\sum_i D_i \left({\bf S}_i \cdot {\bf e}_i\right)^{2} \\ 
- \sum_{ij} K_{ij} \left({\bf S}_i \cdot {\bf S}_j \right)^2 - \sum_i D^{bq}_i \left({\bf S}_i \cdot {\bf e}_i\right)^{4} - 
\sum_{ij} \lambda^{bq}_{ij} (S_i^z S_j^z)^{2}
\end{aligned}
\label{eq:spin-wave1}
\end{equation} 
where $D^{bq}_i$ and $\lambda^{bq}_{ij}$ are the BQ on-site anisotropy and BQ 
anisotropic exchange, respectively. 
We replace the spin operators ${\bf S}_{i,j}$ in Eq.\ref{eq:spin-wave1} 
by bosonic creation ($a^\dagger_i$, $b^\dagger_i$) and 
annihilation ($a_i$, $b_i$) operators over the honeycomb 
sub-lattices $\cal A$ and $\cal B$ using 
Holstein-Primakoff transformations\cite{Holstein40}:  \\

\begin{minipage}{0.49\linewidth} 
\begin{flushright}
\centering
\textbf{For $i\in {\cal A}$ sublattice:}
\begin{eqnarray}
S_i^z & = & (S-a_i^\dagger a_i) \nonumber \\
S_i^+ &\approx & \sqrt{2S} a_i \nonumber \\
S_i^- &\approx & a_i^\dagger\sqrt{2S} \nonumber
\end{eqnarray}
\end{flushright}
\end{minipage}
\begin{minipage}{0.49\linewidth}
\begin{flushleft}
\centering
\textbf{For $i\in {\cal B}$ sublattice:}
\begin{eqnarray}
	S_i^z & = & (S-b_i^\dagger b_i) \nonumber \\
	S_i^+ &\approx & \sqrt{2S} b_i \nonumber \\
	S_i^- &\approx & b_i^\dagger\sqrt{2S}	\nonumber
\end{eqnarray}
\end{flushleft}
\end{minipage}
\\
\\
where we assume that $a_i^\dagger a_i<<S $ and $b_i^\dagger b_i<<S$ 
for small deviations of the spins from their ground state
orientations. This gives (see Supplementary Section \ref{SI-magnontheory} for details): 

\begin{equation}
{\cal \tilde{H}}_{\rm} = \sum_i \left(2\tilde{D} S + Z\cdot S( \tilde{J} + \tilde{\lambda}) \right) ( b^\dagger_i b_i + a^\dagger_i a_i ) - \tilde{J} S  \sum_{\langle ij \rangle} \left( a^\dagger_i b_j +  b^\dagger_j a_i\right)
\label{eq:hamil_mgn}
\end{equation}
where the sum over $i$ and $j$ runs over the sub-lattices and 
first nearest neighbors, respectively, and $Z$ is the number of first 
nearest-neighbors. This procedure outlines a strong implication of the inclusion of 
higher-order exchange interactions in the description of the magnetic 
properties of 2D magnets. That is, the 
enhancement of several magnetic quantities: 
\begin{eqnarray}
\label{J-eq}
\tilde{J}\approx J + 2S^2\cdot K_{\mathrm{bq}}\\
\label{lambda}
\tilde{\lambda}\approx \lambda + 2S^2\cdot \lambda^{\mathrm{bq}}\\
\label{D-eq}
\tilde{D}\approx D + 2S^2\cdot D^{\mathrm{bq}}
\end{eqnarray}
where $\lambda^{\mathrm{bq}}$ and $D^{\mathrm{bq}}$ are 
the BQ anisotropic exchanges, and the BQ on-site magnetic 
anisotropy, respectively, for first nearest neighbors. 
Incidentally Eqs.\ref{J-eq}$-$\ref{D-eq} yield several implications on the magnetic 
properties of the sheets, in particular, on the stabilization of magnetism in truly 2D. 
It is well known 
that in order to overcome thermal fluctuations that could destroy 
any magnetic order\cite{Mermin66}, 
sizable magnetic anisotropies need to be developed to gap the low-energy modes 
in the magnon spectra. That is, a spin wave gap 
needs to appear at the energy dispersion to act as a barrier to 
excitations of long-wavelength spin waves. We can show that the relation between  
the spin wave gap taken into account BQ exchange interactions (${\Delta}_{bq}$) and that 
at the level BL exchange ($\Delta_{bl}$) for first-nearest neighbors 
is given by (see Supplementary Section \ref{SI-magnontheory}):   
\begin{equation}
{\Delta}_{bq} = \Delta_{bl} + 4 S^3 (D^{bq}+\frac{3}{2} \lambda^{bq}) 
\label{spin-gap-BQ}
\end{equation}
where ${\Delta}_{bl} = 2S\left[ {D} + \frac{Z {\lambda}}{2}\right]$. 
By using some parameters for monolayer CrI$_3$
from Table \ref{table2}, and approaching the second term in Eq.\ref{spin-gap-BQ}
as $D^{bq}+3\lambda^{bq}\approx 10.72$ $\mu$eV (see Supplementary Section \ref{SI-sec:BQ-anisotropy})
we can estimate ${\Delta}_{bl}= 0.81$~meV and ${\Delta}_{bq} = 1.0$~meV. 
The magnitude of ${\Delta}_{bq}$ is consistent with recent measurements 
of the magnon dispersion for bulk CrI$_3$, which a spin 
wave gap of approximately 1.3 meV was measured\cite{CrI3_magnons_expermnt}. 
Furthermore, the increment of the on-site and anisotropic 
magnetic anisotropies (Eqs.\ref{lambda}$-$\ref{D-eq})
indicates that not only spin-orbit mechanisms are behind 
the substantial anisotropy in CrI$_3$ but rather 
higher-order exchange processes. 
Such BQ exchange-driven large magnetic 
anisotropy mechanism has been proposed 
for iron-based superconductors\cite{BQpictinides,Iron-BQ,unifiedBQ} 
which successfully described their magnetic properties.  
A direct consequence Eqs.\ref{J-eq}$-$\ref{D-eq} is the increment of Curie temperatures 
by factor $r$ given by (see Supplementary Section \ref{SI-magnontheory}): 
\begin{equation}
r=\frac{\tilde{T}_C}{T_C} \approx \frac{\tilde{J}  \ln (1 + 2\pi J S / \Delta_{bl})}{J  \ln(1 + 2\pi\tilde{J}S / {\Delta}_{bq})}.
\label{r-factor}
\end{equation}
where $\tilde{T}_C$ and $T_C$ are the 
Curie temperatures with and without BQ interactions, respectively. 
Including few values in Eq.\ref{r-factor}, we can roughly estimate an 
enhancement of $r\approx39\%$ for monolayer CrI$_3$ which follows 
the Monte Carlo calculations (Fig.~\ref{fig3}{\bf a-b}).  
It is worth noticing that the model in Eq.~\ref{eq:hamil_mgn},
{\it i}) takes into account only first-nearest neighbors in the exchange interactions, and 
{\it ii}) we assume a mean-field approach 
in the solution of the non-linear Holstein-Primakoff transformation (e.g. magnon-magnon interactions)
to simplify the complex mathematical terms, i.e. four-operator product. 
Supplementary Section \ref{SI-four-operator} provides a full discussion on the details involved. 


\section{Interplay between biquadratic exchange and Dzyaloshinskii-Moriya interactions in topological spin excitations}


An intriguing question that raised by the presence of BQ exchange interactions is 
whether they play an important role in the description of magnetic quasiparticles such as magnons and 
non-trivial spin textures in 2D vdw magnets. It has recently been shown using 
neutron scattering\cite{CrI3_magnons_expermnt} that CrI$_3$ magnet
shows topological spin-excitations with two distinctive magnon bands 
separated by a bandgap of 4~meV at the Dirac {\bf K}-point. 
In spite of the clear demonstration that CrI$_3$ can not follow an Ising model as initially 
pointed out\cite{firstCrI3}, these results indicate that non-Heisenberg interactions play an 
important role in the creation of spin-excitations in 2D magnetic materials. 
Since the gap opening at {\bf K} is related with the inversion symmetry 
breaking and appearance of DMI, chirality becomes crucial in the discrimination 
of the magnon bound states. 
Moreover, it has become well established\cite{Elliott69, Castets72,Fisk01,BQpictinides} that 
isotropic spin interactions at the level of the BL Heisenberg models do not capture all features 
in the energy dispersion of spin-excitations in magnetic materials. 
There are additional contributions through uniaxial anisotropies, next-nearest neighbor interactions and  
the delicate balance between them, that need to be considered. In order to account for all these quantities, 
we extended the model in Eq.\ref{eq:hamil_bq} with the addition of DMI: 
\begin{equation}
{\cal H}_{\rm latt}= {\cal {H}}_{\rm } + \sum_{\left<\left<ij\right>\right>} \bm{A}_{ij} \cdot \left( {\bf S}_i \times {\bf S}_j \right)
\label{bq_dmi}
\end{equation}
where $\bm{A}_{ij}$ is the DMI between spins $\bm{S}_{i}$ and $\bm{S}_{j}$. For a honeycomb 
ferromagnetic layer, with an easy axis perpendicular to the surface ({\bf z}-direction), 
there is no breaking of the inversion symmetry 
of the lattice at first nearest-neighbors, e.g. $\bm{A}_{ij}^{1st} = 0$. However, 
contributions from the second nearest-neighbors become non-negligible as space 
inversion is not present. Therefore, we consider the DMI vector as 
$\bm{A} = \nu_{ij}A^z\bm{z}$, where $A_z$ is the magnitude of the 
DMI along of the easy-axis, and $\nu_{ij}=\pm 1$ represents the hopping of spins at second nearest-neighbors
from sites $i$ to $j$ and vice-versa, respectively (Fig.~\ref{fig2}{\bf a}). 
Similarly as shown above,  
we can use Holstein-Primakoff transformations for $J_i > 0$ to write Eq.\ref{bq_dmi}  
in terms of bosonic creation and annihilation operators (see details in Supplementary Section \ref{SI-dmi+bq}): 
\begin{equation}
 {\cal H}_{latt} =  {\cal H}_{BL} + {\cal H}_{BQ} + {\cal H}_{DMI}  
\label{full-H}
\end{equation}
where we separate the terms due to BL exchange (${\cal H}_{BL}$),  
from those due to BQ (${\cal H}_{BQ}$) and DMI (${\cal H}_{DMI}$) 
which can be written as: 

\begin{equation}
{\cal H}_{BL} = H_D + \sum_{i=1}^{3} H_i^{Isotropic} +  \sum_{i=1}^{3} H_i^{Anisotropic} 
\label{BL-ham}
\end{equation}
\begin{equation}
H_{BQ}  =  -12K_{bq}S^4 -2K_{bq} S^3 \left( \sum_{\left<ij\right>} \left( b_j^\dagger a_i + a_i^\dagger b_j \right)  - Z_1 \sum_{i \in {\cal A}}^{N/2} a_i^\dagger a_i  -Z_1  \sum_{i \in {\cal B}}^{N/2} b_i^\dagger b_i \right)
\label{BQ-ham}
\end{equation}
\begin{equation}
{\cal H}_{DMI} = iA_zS\left( \sum_{\left<\left<ij\right>\right>\in {\cal A}}^{N/2} \left( a_i^\dagger a_j - a_j^\dagger a_i\right) + \sum_{\left<\left<ij\right>\right>\in {\cal B}}^{N/2} \left( b_i^\dagger b_j - b_j^\dagger b_i\right)  \right)
\label{DMI-ham}
\end{equation}
where $H_D$ is the on-site anisotropy term, 
and $H_i^{Isotropic}$ and $H_i^{Anisotropic}$ are respectively
the isotropic and anisotropic parts of the BL exchange Hamiltonian 
taken into account up to third nearest-neighbors ($i=1,2,3$). 
The sum in $\left<ij\right>$ runs over the first nearest-neighbors at 
both sublattices $\cal A$ and $\cal B$ while that on $\left<\left<ij\right>\right>$ 
runs over the second nearest-neighbors
specifically on either $\cal A$ or $\cal B$ lattice. We notice that 
the second nearest neighbors not only break the inversion symmetry 
of the honeycomb lattice but also generate a magnetic flux $\phi$ 
involving $A_z$ and $J_2$  given by: 

\begin{equation}
\phi = tan^{-1} (A_z/J_2)
\label{flux}
\end{equation}
The magnetic flux (circular arrow) can be appreciated in Fig.~\ref{fig2}{\bf a} 
when the magnons (dashed lines) hop between the second nearest-neighbors.  
This process introduces a phase $\phi_{ij}=\mu_{ij}\phi$ ($\mu_{ij}=\pm 1$) 
in the magnons as they hop from a site $i$ to $j$, and vice-versa. 
The different magnitudes of $\mu_{ij}$ determine whether the hopping follows the flux 
and consequently induces nontrivial topological properties (e.g. chirality) similarly as in the 
Haldane model\cite{Haldane88,Owerre16} for fermions. 
The topological features of the magnon bands can be described in the ${\bm k}-$space by using 
the Fourier transform of the creation ($a_{\bm{k}}^\dagger$, $b_{\bm{k}}^\dagger$) 
and annihilation ($a_{\bm{k}}$, $b_{\bm{k}}$) operators in Eq.\ref{full-H} as: 
\begin{equation}
    \mathcal{H} = {\cal H}_0 + \sum_{\bm{k}} \begin{pmatrix}
a^\dagger_{\bm{k}} & b^\dagger_{\bm{k}} 
\end{pmatrix}
\begin{pmatrix}
h_0(\bm{k}) + h_z(\bm{k})  & h_x(\bm{k}) - ih_y(\bm{k}) \\
h_x(\bm{k}) + ih_y(\bm{k}) & h_0(\bm{k}) - h_z(\bm{k}) \\
\end{pmatrix}
\begin{pmatrix}
a_{\bm{k}} \\
b_{\bm{k}}
\end{pmatrix}
\label{matrix}
\end{equation}
where the different terms can be written as (see Supplementary Section \ref{SI-dmi+bq} for details): 
\begin{eqnarray}
    {\cal H}_0 & = &-2DS -3(J_1+\lambda_1)S^2 - 6(J_2+\lambda_2)S^2 - 12KS^4 \nonumber \\
    h_0(\bm{k}) & = &  \varepsilon_0 - 4\sqrt{(J_2S)^2 + (A_zS)^2}C(\bm{k})  \nonumber    \\
    \varepsilon_0 & = & 2DS + (J_1+\lambda_1)Z_1S + 6KS^3 + 6(J_2+\lambda_2)S +3J_3S   \nonumber \\
    h_x(\bm{k}) & = & -(J_1 + 2KS^2)S \sum_{j=1}^3 \mathrm{cos}(\bm{k}\cdot \bm{\tau_j}) - J_3S\sum_{j=1}^3 \mathrm{cos}(\bm{k}\cdot (\bm{u_j} +\bm{\tau_j}))   \nonumber   \\
    h_y(\bm{k}) & = & -(J_1 + 2KS^2)S \sum_{j=1}^3 \mathrm{sin}(\bm{k}\cdot \bm{\tau_j}) -J_3S\sum_{j=1}^3 \mathrm{sin}(\bm{k}\cdot (\bm{u_j} +\bm{\tau_j}))  \nonumber   \\
    h_z(\bm{k}) & = & 4\sqrt{(J_2S)^2 + (A_zS)^2}S(\bm{k})  
    \label{many-eq}
\end{eqnarray}  
where $C(\bm{k})=  \mathrm{cos}(\phi)\sum_{j=1}^3\mathrm{cos}(\bm{k}\cdot\bm{u_j})$ and 
$S(\bm{k}) = \mathrm{sin}(\phi)\sum_{j=1}^3\mathrm{sin}(\bm{k}\cdot\bm{u_j}) $. 
The vectors $\bm{\tau_j}$ and $\bm{u_j}$ are respectively 
between $1^{st}$ and $2^{nd}$ nearest neighbors (Fig.~\ref{fig2}{\bf a}).

The eigenvalues of Eq.\ref{matrix} can be written in terms of 
the lower and upper energy bands as: 
\begin{equation}
    E_{}^{\pm} = h_0(\bm{k}) \pm \sqrt{h_x(\bm{k})^2 + h_y(\bm{k})^2 + h_z(\bm{k})^2}  
    \label{dispersion}
\end{equation}
Note that Eqs.\ref{many-eq} and Eq.\ref{dispersion} are general for any 
honeycomb material with ferromagnetic order, easy-axis perpendicular to 
the surface and develop DMI and BQ exchange interactions. For instance, we can use them to predict 
the energy dispersion of the magnon bands over the first Brillouin zone (BZ) of 
any 2D magnet, e.g. CrI$_3$. Figure \ref{fig4}{\bf a-e} shows 
different levels of theory either using a simple XXZ model or including more sophisticated terms 
through the BQ exchange, DMI or simultaneously all of them. It is clear that XXZ models without any 
contribution from DMI (Fig.~\ref{fig4}{\bf a-c}) does not describe the gap opening at the Dirac points due to the breaking of the inversion symmetry. 
Moreover, a model at the level of XXZ$+$DMI as initially used to understand the 
magnon dispersion of CrI$_3$\cite{CrI3_magnons_expermnt}
does not capture entirely 
the full profile of the bands (Fig.~\ref{fig4}{\bf d}). 
The upper branch $E_{}^{+}$ becomes nearly flat with 
the increment of $J_2$ at the path $K-M-K$ while the lower magnon branch 
$E_{}^{-}$ turns more curved. This picture modifies substantially when BQ exchange 
interactions are included in the XXZ$+$BQ$+$DMI model 
(Fig.~\ref{fig4}{\bf e}) as the magnon bands follow a 
similar curvature throughout the variation of $J_2$, although the upper 
branch at $\Gamma$ increased to $E_{\Gamma}^{+}=$27.8 meV. 
A sound comparison between the XXZ$+$BQ$+$DMI model and the experimental 
results for CrI$_3$ is obtained (Fig.~\ref{fig4}{\bf f}) when $J_1$ is 
varied similarly (Fig.~\ref{fig4}{\bf a}) indicating that the first 
nearest-neighbors are important on the stabilization of $E_{\Gamma}^{+}$. 
It is worth mentioning that the value of the exchange 
interactions taken into account in the fitting of the neutron scattering 
spectra\cite{CrI3_magnons_expermnt} do not separate BL contributions 
from BQ as shown in Eq.\ref{J-eq}-\ref{D-eq}. Hence, it is not known from 
the fitting procedure\cite{CrI3_magnons_expermnt} what is 
the contribution of $K_{bq}$ to the magnon dispersion. 
However, such separation can be clearly stated in our model as 
indicated in Fig.~\ref{fig4}{\bf f}. Furthermore, 
even though DMI is important for the gap opening at the 
Dirac point, it does not contribute 
to the magnitudes of the magnetization or critical temperatures for any 2D magnet
with an out-of-plane easy-axis and ferromagnetic aligned spins (see details in 
Supplementary Section \ref{SI-dmi+bq}).

\section{Implications and prospects}  

The effects of the BQ exchange interactions proposed here should manifest 
in experimentally accessible temperature range. One indication is already 
the accurate reproduction of experimental Curie temperatures\cite{firstCrI3,kostya19} 
including higher-order exchange which could not be obtained at the level of 
Ising, Heisenberg or Kitaev models. The magnitudes of BQ exchange can be in principle extracted from 
accurate hysteresis loops using phenomenological models\cite{Jonge00,Bader96}. 
Importantly in such analyses are potential temperature variations of BL and BQ exchanges with 
layer thickness which may indicate tunable interlayer exchanges still to be explored in 2D magnets. 

Frustrated 2D Heisenberg models in the presence of BQ exchange interactions 
are also a non-trivial matter with a rich phase diagram involving 
incommensurate spin spirals, canted ferromagnetic states, quadrupolar phase or vertexes\cite{Mahanti10,frustrated, quadrupolarphase,BQpictinides}. As our results indicate that BQ exchanges 
are important for 2D magnets, possible ordered and disordered magnetic 
regimes may be stabilized in appreciable temperatures. 
Moreover, it is possible to enhance or suppress BQ interactions in Mott-Hubbard systems by 
applying external electric fields\cite{Katsnelson19}. Indeed, 
the control of the magnetic properties of CrI$_3$ using electrical 
means has already been demonstrated\cite{2LCrI3}. Therefore, 
this opens the prospect of a coherent transfer between 
spin and charge degrees of freedom using short laser pulses in nanosheets. 

In summary, we have shown the importance of biquadratic exchange interaction in the magnetic 
properties of 2D materials. We have described the phenomenology of such higher-order spin coupling, discussed its 
implications on several magnetic properties, and presented results at the level of non-collinear 
first-principles methods, Monte Carlo approximations and analytical models. 
The developed spin Hamiltonian including BQ exchange and DMI provided an 
accurate picture of topological spin-excitations on a generalized basis for any 2D magnet. 
Our results are particularly timely given the increasing interest in quantum materials, 
and we believe that our work will motivate the exploration of different exchange 
couplings and competition between critical phenomena\cite{Basov17}.

\section{Acknowledgements}

EJGS acknowledges computational resources through the 
UK Materials and Molecular Modeling Hub for access to THOMAS supercluster, 
which is partially funded by EPSRC (EP/P020194/1); and CIRRUS Tier-2 HPC 
Service (ec019 Cirrus Project) at EPCC (http://www.cirrus.ac.uk) funded 
by the University of Edinburgh and EPSRC (EP/P020267/1). 
EJGS acknowledges the Department for the Economy (USI 097), 
EPSRC Early Career Fellowship (EP/T021578/1) and 
the University of Edinburgh for funding support.

\section*{Supplementary Materials}
\label{sec:org3881bef}

Supplementary sections

Materials and Methods.

\subsubsection{Data Availability}

The data that support the findings of this study 
are available within the paper and its Supplementary Information.  

\subsubsection{Competing interests}
The Authors declare no conflict of interests.

\subsubsection{Author Contributions}
EJGS conceived the idea and supervised the project. 
AK performed the first-principles and Monte Carlo 
simulations under the supervision of EJGS. 
RFLE implemented the biquadratic exchange interactions in 
VAMPIRE. MA, AK and EJGS developed the analytical and 
numerical models used to describe the biquadratic exchange and DMI. 
KSN contributed on the discussions and significance of the results. 
EJGS wrote the paper with inputs from all authors. 
All authors contributed to this work, 
read the manuscript, discussed the results, and agreed 
to the contents of the manuscript.

\pagebreak

\bibliography{ref_bq}

\pagebreak 

\begin{table}
\begin{tabular}{c|c|c|c} 
		\hline
		\rowcolor{gray!15}
     Material & $\mu_M$  ($\mu_B$) & S & $K_{\mathrm{bq}}$ (meV) \\
     \hline\hline 
     \rowcolor{blue!23}
	CrI$_3$      &  3.36  & 1.5 & 0.21  \\
	\rowcolor{blue!23}
	CrBr$_3$     & 3.17  & 1.5 & 0.22  \\
	\rowcolor{blue!23}
	CrCl$_3$     & 3.05  & 1.5 & 0.21 \\
	\rowcolor{blue!23}
		CrF$_3$      & 2.94  & 1.5 & 0.09 \\
	\rowcolor{blue!23}
		MnBr$_3$     & 3.88  & 2   & 0.04 \\
     \rowcolor{blue!23}
		CuBr$_3$     & 0.53  & 0.5 & 39.16  \\
   \rowcolor{blue!23}
		FeBr$_3$     & 3.88  & 2   & 0.06  \\
   \rowcolor{blue!23}
		VBr$_3$      & 1.93  & 1   & 2.85  \\
   \rowcolor{blue!23}
		CrGeTe$_3$ & 2.81 & 1.5 & 0.35 \\
		   \rowcolor{blue!23}
		CrPTe$_3$  & 2.92  & 1.5 & 0.35 \\
		   \rowcolor{blue!23}
		CrSiTe$_3$ & 3.13 & 1.5 & 0.31 \\
		   \rowcolor{blue!23}
		MnPS$_3$   & 4.24  & 2   & 0.02 \\
		   \rowcolor{blue!23}
		MnPSe$_3$  & 4.21  & 2   & 0.02 \\
		   \rowcolor{blue!23}
		MnPTe$_3$  & 3.93 & 2   & 0.11  \\
		   \rowcolor{yellow!23}
		1T-MnSe$_2$   & 3.49  & 1.5 & 0.74 \\
		  \rowcolor{yellow!23}
		2H-MnSe$_2$  & 3.49  & 2   & 0.15 \\
		  \rowcolor{yellow!23}
		1T-MnTe$_2$  & 3.71  & 2    & 0.14 \\
		  \rowcolor{yellow!23}
		2H-MnTe$_2$  & 3.78  & 2    & 0.10 \\
		  \rowcolor{yellow!23}
		1T-MnS$_2$    & 3.28  & 1.5 & 1.09 \\
		  \rowcolor{yellow!23}
		2H-MnS$_2$   & 3.27  & 1.5 & 1.55  \\
		  \rowcolor{yellow!23}
		2H-FeS$_2$    & 2.18  & 1    & 20.11 \\
		\hline
	\end{tabular}
\caption{Calculated biquadratic exchange ($K_{bq}$) for several 2D magnetic materials with honeycomb (faint blue) and hexagonal (faint red) lattices using non-collinear ab initio methods as explained in the text. The magnitudes of the spin
angular momentum ($S$) used in the model (Supplementary Section \ref{SI-BQcalc}) 
and the magnetic moments at the metal atoms ($\mu_M$) are also included. 
}	\label{table1}
\end{table}

\newpage

\begin{table}
	\small
	\begin{tabular}{|c|c|c|c|c|c|c|c|c|} 
\hline
\cellcolor{gray!15}Comp. & \cellcolor{gray!15}$ \cellcolor{gray!15}J_{1}$ (meV) & \cellcolor{gray!15}$J_{2}$ ($\mu$eV)  & \cellcolor{gray!15}$J_{3}$ ($\mu$eV)  & \cellcolor{gray!15}$\lambda_{1}$ ($\mu$eV) & \cellcolor{gray!15}$\lambda_{2}$ ($\mu$eV)&$ \cellcolor{gray!15}\lambda_{3}$ ($\mu$eV) & \cellcolor{gray!15}$D$ ($\mu$eV) \\
		\hline\hline
		   \rowcolor{blue!23}
		CrI$_{3}$   & 2.01 & 320.02 & 8.10 & 106.8 & -10.24 & 0.91 & 108.82 \\

		   \rowcolor{blue!23}
		CrBr$_{3}$  & 1.66 & 164.35 &-11.60 & 20.69 & -2.06 & -0.69 & 34.09 \\ 
		   \rowcolor{blue!23}
		CrCl$_{3}$  & 1.28 & 72.03 & -25.18 & 20.07 & -9.74 & -0.51 & 12.67 \\
		   \rowcolor{blue!23}
		CrF$_{3}$   &-0.23 & 17.27 & 0.20 & 3.33 & -0.67 & -0.14 & 122.02 \\
		\hline
	\end{tabular}
	\caption{Computed values of several magnetic quantities for CrX$_3$ (X=I, Br, Cl, F) at different number of nearest neighbors: isotropic ($J_1$, $J_2$, $J_3$) and anisotropic ($\lambda_1$, $\lambda_2$, $\lambda_3$) BL exchanges. 
	The on-site magnetic anisotropy $D$ is also included. See Supplementary Section \ref{SI-NNN_exch} for details. 
	}
	\label{table2}
\end{table}

\newpage
%

%
\begin{figure}
	\includegraphics[width=1.0\textwidth]{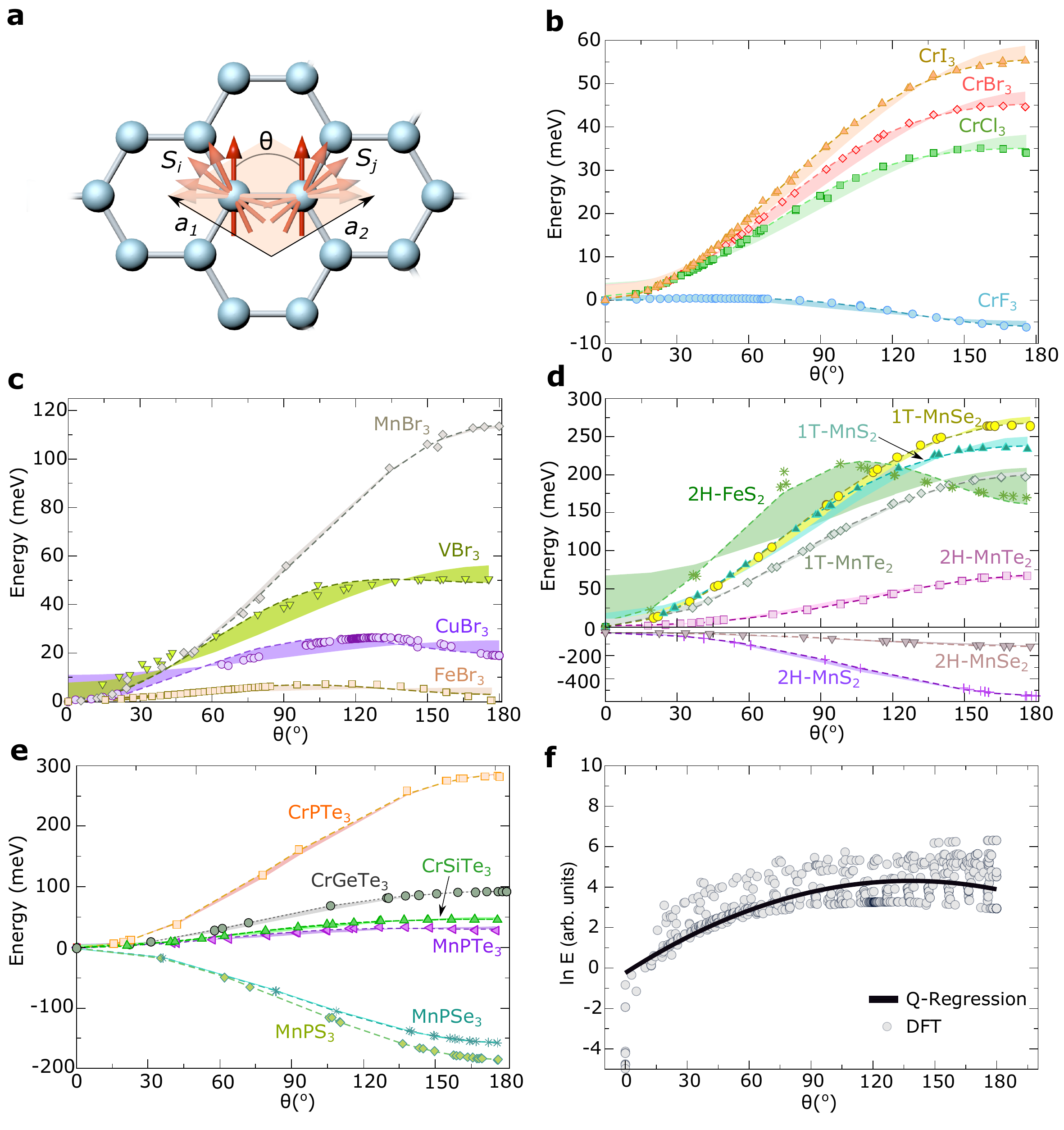}
	\caption{ {\bf }  
\label{fig:etot_rotscheme}}
\end{figure}


%
\begin{figure}
\includegraphics[width=0.75\textwidth]{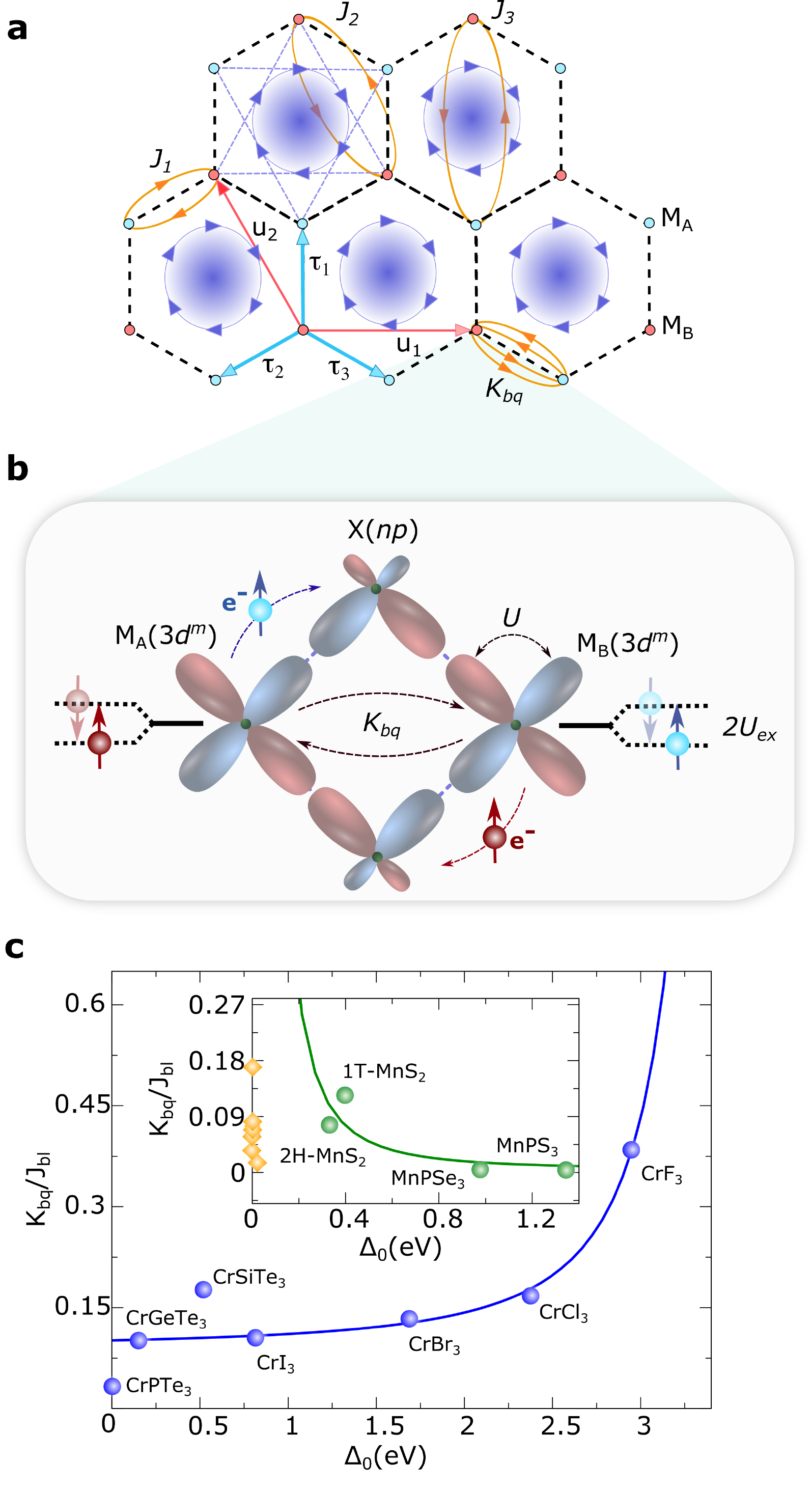} 
\caption{
{\bf }  
\label{fig2}}
\end{figure}


%
\begin{figure*}
	\includegraphics[width=1.05\textwidth]{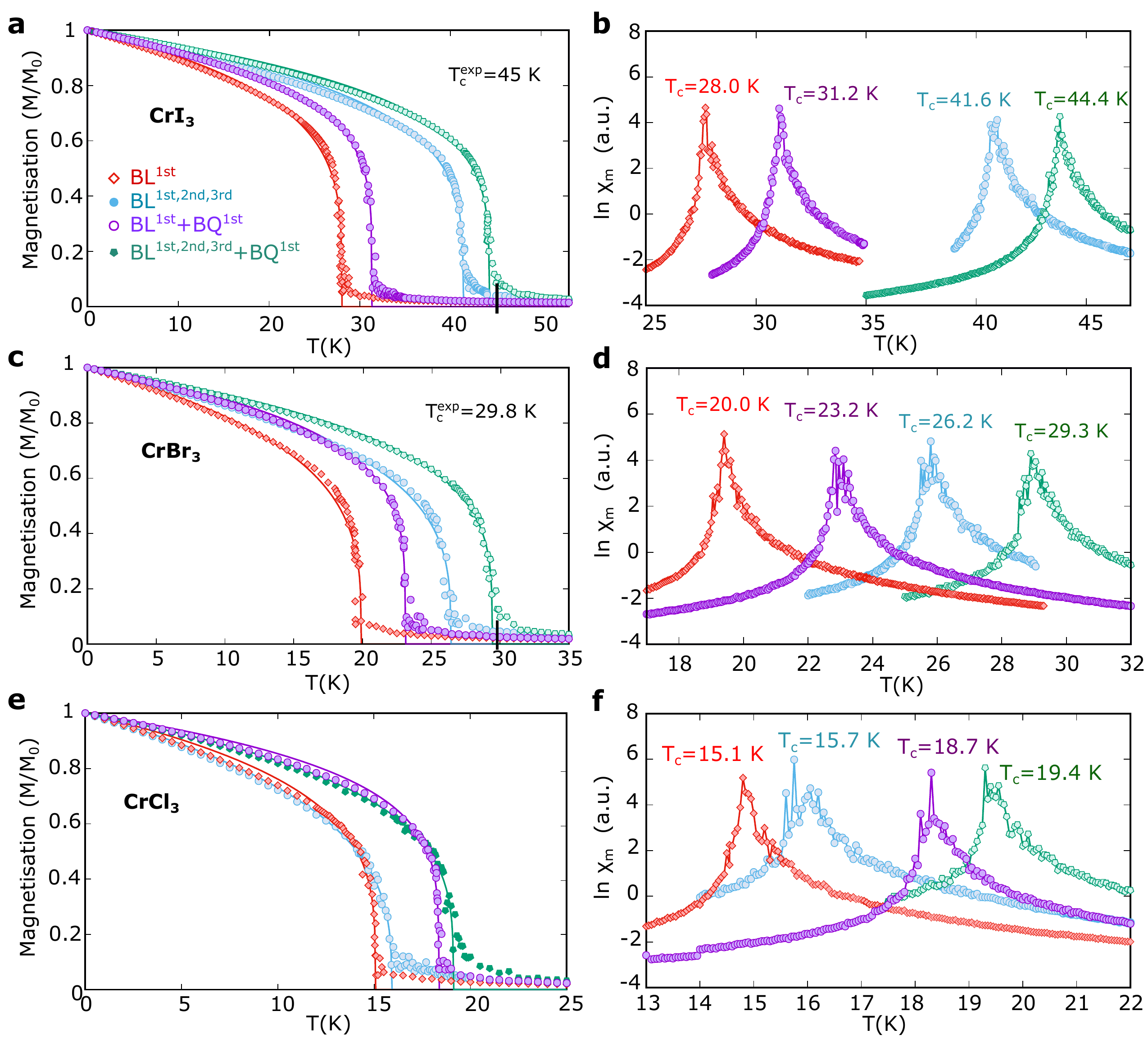}
	\caption{{\bf }  	
\label{fig3}
}
\end{figure*}

\pagebreak

\begin{figure}
	\includegraphics[width=\textwidth]{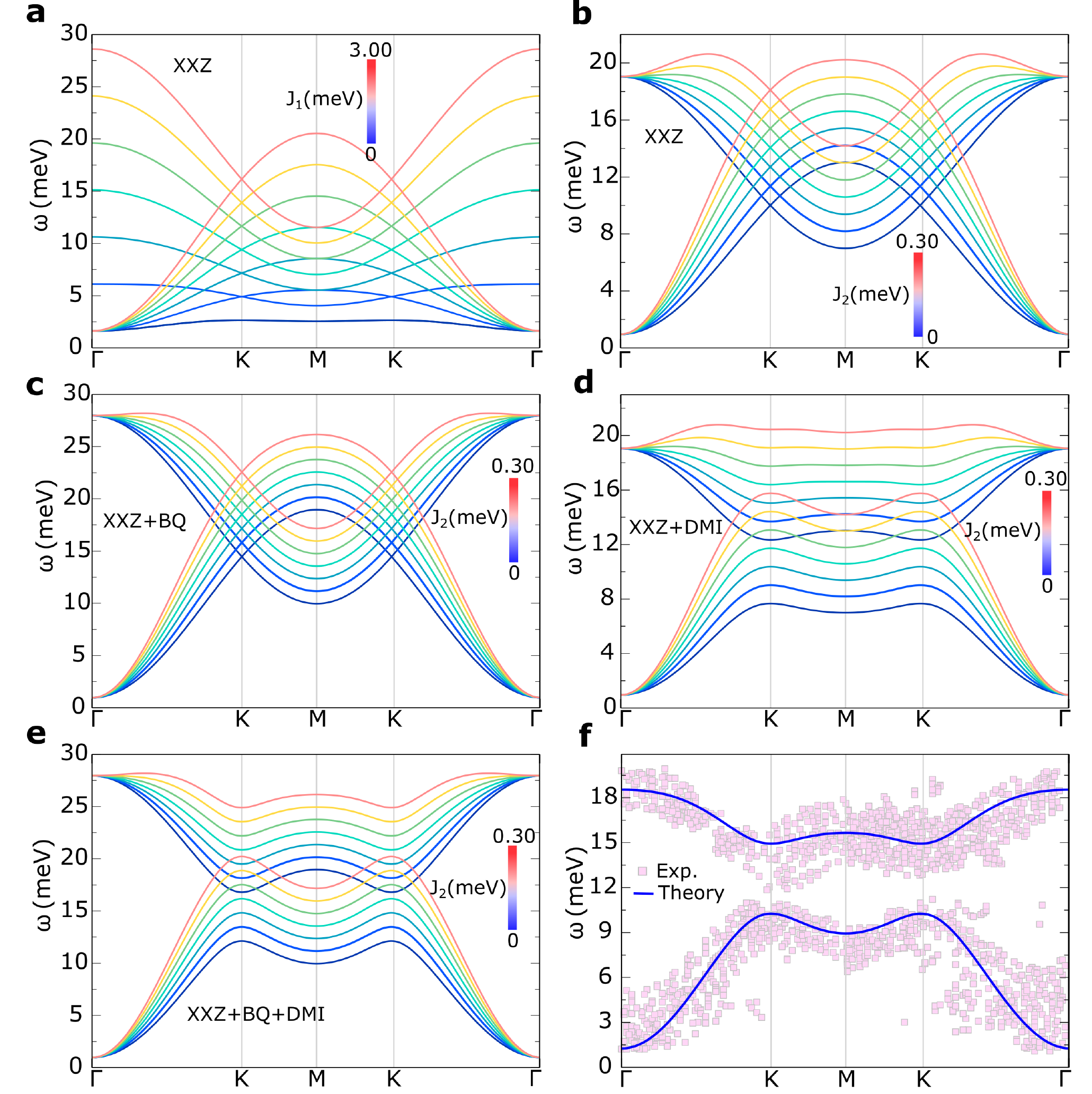}
	\caption{{\bf} 
\label{fig4}	
}
\end{figure}

\pagebreak

\setcounter{figure}{0}


%
\begin{figure}
	\caption{ {\bf Biquadratic exchange interactions in 2D magnets.}
{\bf a,} Diagram of the rotation of spins $S_i$ and $S_j$ in the unit cell (defined by vectors {\bf a}$_1$ and {\bf a}$_2$) 
of a 2D magnet by a relative angle $\theta$ between them. 
Spins are rotated symmetrically 
in opposite directions from 0$^{\rm o}$ to 180$^{\rm o}$. 
{\bf b-e,} Relative total energy (eV) as a function of $\theta$($^{\rm o}$) for different monolayers of 2D magnets: 
trihalides (CrX$_3$, X=F, Cl, Br, I), metal tribromides (MBr$_3$, M=Mn, Cu, Fe, V), 
chromium based ternary tellurides (Cr$_{\rm 2}$X$_{\rm 2}$Te$_{\rm 6}$, X=Ge, P, Si), 
manganese based ternary chalcogenides (Mn$_{\rm 2}$P$_{\rm 2}$X$_{\rm 6}$, X=S, Se, Te), 
transition metal dichalcogenides (MnX$_{\rm 2}$, X=S, Se, Te) of different phases ($2H$, $1T$) 
and an iron-based dichalcogenide (2H-FeS$_{\rm 2}$).  The reference energy is taken at 
0$^{\rm o}$ as the spins oriented at the same direction. Rotations can occur in-plane or out-of-plane 
with similar behavior. Symbols are calculated energies. Dashed lines correspond to a 
quadratic fitting using $E^{tot}_{bq}\left(\theta \right)=A^{bq}_0+A^{bq}_1\cdot S^2 \cos(\theta)+A^{bq}_2\cdot S^4 \cos^2(\theta)$, 
while color filling areas indicate the deviation between the quadratic $E_{bq}^{tot}$ 
and a linear fitting using $E_{bl}^{tot}\left(\theta\right)=A_0^{bl}+A_1^{bl}\cdot S^2 \cos(\theta)$. 
Materials that show large deviation, such as CuBr$_3$ or 2H-FeS$_2$, develop large BQ exchange interactions.  
{\bf f,} Logarithm of the total energies for the dataset in {\bf b-e,} as a function of $\theta$ (dots). A quadratic regression 
(Q-Regression) is evaluated over the calculated DFT energies (solid line) indicating a universal behavior of BQ exchange interactions in 2D magnets.   
\label{fig:etot_rotscheme}}
\end{figure}

\pagebreak

\begin{figure}
\caption{
{\bf Bilinear (BL) and biquadratic (BQ) exchange interactions in 2D magnets.} 
{\bf a,} Schematic of the BL exchanges at first (J$_{\rm 1}$), 
second (J$_{\rm 2}$) and third (J$_{\rm 3}$) nearest neighbors (NN), and BQ exchange 
($K_{bq}$) at first NN. Single and double line diagrams represent 
BL and BQ exchanges, respectively. 
The two inequivalent magnetic sites in the honeycomb lattice 
are shown by faint blue (M$_{\rm A}$) and faint red (M$_{\rm B}$) dots. 
The blue orbits inside of the hexagons represent the magnetic flux $\phi$ generated 
by the second-NN Dzyaloshinskii-Moriya 
interactions (DMI), which breaks the inversion symmetry of the lattice. 
The dashed lines show the magnon hopping between second NN as the magnons 
gain a phase given by $\phi$ (see text). The lattice 
vectors {\bf u}$_{i}$ ($i=1, 2$) and $\tau_{j}$ ($i=1, 2, 3$) 
show the first and second NN on the lattice, respectively. 
{\bf b,} Zoom-in on the BQ exchange process involving two electrons between sites 
M$_{\rm A}$ and M$_{\rm B}$ with 3$d^m$ electrons in the valence. The BQ exchange $K_{bq}$
is mediated by non-magnetic atoms X with a valence given by $np$ electrons, 
where $n$ will depend on the atomic elements involved. An on-site Hubbard $U$ 
term is at the M$_{\rm A,B}$ sites with $U_{ex}$ representing the potential 
spin-splitting when the spins of the electrons involved in the BQ exchange 
align ferro- or anti-ferromagnetically (Supplementary Section \ref{SI-sec:fm-afm_competition}). 
The difference between up and down spin coupling is given by $2U_{ex}$. 
{\bf c,} K$_{bq}$/J$_{bl}$ versus $\Delta_{0}$(eV) for all materials 
displaying BQ exchange interactions. Magnetic atoms with 
similar chemical environment in terms of Coulomb repulsion, 
exchange interactions and valence follow alike 
behavior for K$_{bq}$/J$_{bl}$. For instance, Cr in blue and 
Mn in green (inset). Orange dots show materials with dissimilar 
electronic configurations but with $\Delta_{0}=0$. 
\label{fig2}}
\end{figure}


%
\begin{figure*}
	\caption{{\bf Monte-Carlo simulations at different levels of theory 
	including BL and BQ exchange interactions.}  
{\bf a-b} Magnetization (M/M$_{\rm 0}$) and logarithm of the magnetic longitudinal 
susceptibility ln$\chi_m$ (a.u.) versus temperature (K), respectively, for monolayer CrI$_{\rm 3}$. 
Calculated and fitting curves using M/M$_{\rm 0}= (1-{\rm T/T}_{\rm c}$)$^{\beta}$ 
(where T$_{\rm c}$ and $\beta$ are the critical temperature and coefficient, respectively) are shown by 
dots and solid lines, respectively, in {\bf a}. Solid lines in {\bf b} show the interpolation between points.
Different curves correspond to different number of  NN, from one up to third, 
taken into account in BL interactions: BL$^{\rm 1st}$ (faint red), BL$^{\rm 1st, 2nd, 3rd}$ (faint blue). 
Results including BQ exchange at first NN (BQ$^{\rm 1st}$) with different number of BL 
exchanges are shown in purple (BL$^{\rm 1st}+$BQ$^{\rm 1st}$) and 
faint green (BL$^{\rm 1st, 2nd, 3rd}+${\rm BQ}$^{\rm 1st}$). 
Critical temperatures (T$_{\rm c}$) at each level of BL and BQ exchange interactions 
are indicated at {\bf b} with the maximum magnitude of ln$\chi_m$ (a.u.) highlighted. 
Magnetic susceptibility is shown in logarithm scale for clarify.  Experimental critical 
temperature (T$_{\rm c}^{\rm exp}$) is included for comparison. 
{\bf c-d,} and {\bf e-f,} similar plots as in {\bf a-b,} for CrBr$_{3}$ and CrCl$_{3}$, respectively. 
Critical exponents $\beta$ extracted from the simulations 
for BL$^{\rm 1st, 2nd, 3rd}+{\rm BQ}^{\rm 1st}$ 
are $\beta=$0.22, 0.24 and 0.28 for CrI$_3$, CrBr$_3$ and CrCl$_3$ respectively. 
Magnitudes of $\beta$ for BL$^{\rm 1st, 2nd, 3rd}$ are 0.25, 0.28 and 0.32 for CrI$_3$, 
CrBr$_3$ and CrCl$_3$ respectively. 	
\label{fig3}
}
\end{figure*}

\pagebreak

\begin{figure}
	\caption{{\bf Magnon spectra ${\bm \omega}$(meV) at different levels of theory for monolayer CrI$_{\rm 3}$.} 
{\bf a,} $\omega$(meV) versus {\bf k} at the first Brillouin zone ($\Gamma - K- M - K - \Gamma$) 
using a XXZ model. $J_{\rm 1}$ varies within $0 - 3.50$ meV in steps 
of 0.5 meV from each curve (color map). 
{\bf b,} Similar as {\bf a}, but with a fixed $J_{\rm 1}=2.01$ meV and varying $J_{\rm 2}$ 
within $0 - 0.30$ meV (color map) in steps of 0.05 meV. 
{\bf c-e,} $\omega$(meV) versus {\bf k} for different models: XXZ including BQ exchange (XXZ+BQ), XXZ including DMI (XXZ+DMI), and XXZ including both BQ exchange and DMI (XXX+BQ+DMI). In these plots, 
$J_{\rm 1}=2.01$ meV, $A_z=0.31$ meV\cite{CrI3_magnons_expermnt} 
(on {\bf d} and {\bf e}) and $J_{\rm 2}$ varies within $0 - 0.30$ meV. 
{\bf f,} Comparison between the XXZ+BQ+DMI model and the experimental data\cite{CrI3_magnons_expermnt}
recently measured for bulk CrI$_3$. We used as parameters:    
$J_{\rm 1}=1.01$ meV, $J_{\rm 2}=0.10$ meV, $K_{bq}=0.22$ meV.
\label{fig4}}
\end{figure}

\clearpage

\end{document}